\documentclass{article}
\begin{document}

\vspace{1cm}
\begin{flushright}
CAMS/00-10
\end{flushright}
\vspace{1cm} \baselineskip=16pt

\begin{center}
\baselineskip=16pt \centerline{{\Large{\bf Complex Gravity and Noncommutative
Geometry \footnote{Talk given at the String 2000 meeting, July 10-15 2000,
University of Michigan, Ann Arbor, USA} }}} \vskip1 cm

Ali H. Chamseddine  \vskip1cm \centerline{\em Center for Advanced Mathematical
Sciences (CAMS) } \centerline{\em and} \centerline{\em Physics Department,
American University of Beirut, Lebanon}
\end{center}

\vskip1 cm \centerline{\bf ABSTRACT} The presence of a constant background
antisymmetric tensor for open strings or D-branes forces the space-time
coordinates to be noncommutative. An immediate consequence of this is that
all fields get complexified. By applying this idea to gravity one discovers
that the metric becomes complex. Complex gravity is constructed by gauging
the symmetry $U(1,D-1)$. The resulting action gives one specific form of
nonsymmetric gravity. In contrast to other theories of nonsymmetric gravity
the action is both unique and gauge invariant. It is argued that for this
theory to be consistent one must prove the existence of generalized
diffeomorphism invariance. The results are easily generalized to
noncommutative spaces. \vfill\eject
\bigskip

At planckian energies, the manifold structure of space-time will not hold,
and a new geometrical setting is needed. At present there are two\ possible
candidates to describe space-time at high energies, one is string theory and
the other is noncommutative geometry \cite{CF}. Recently these two
approaches came togother when it was realized that the presence of constant
background $B$-field for open strings or D-branes implies that the
coordinates of space-time become noncommuting (\cite{CDS},\cite{DH},\cite{CK}%
, \cite{CH},\cite{S},\cite{AFS},\cite{SW}). This result is expected to
generalize to the case of a non-constant $B$-field. The resulting
geometrical space is expected to be noncommuting and curved. The question I\
will address in this talk is how to describe the dynamics of the
gravitational field in such spaces.

\bigskip One possibility is to use the tools of noncommutative geometry of
Alain Connes as specified by the spectral triple $\left( \mathcal{A},%
\mathcal{H},D\right) $ where $\mathcal{A}$ is an associative algebra with a
* product and identity, $\mathcal{H}$ a Hilbert space and $D$ a self-adjoint
operator on $\mathcal{H}$ such that $\left[ D,a\right] $, $a\in \mathcal{A}$
defines a bounded operator on $\mathcal{H}$ \cite{bconnes}. In this setting
it is possible to develop the noncommutative analogue of Riemannian
geometry. A good example of the realization of noncommutative geometry is
the data encoded in superconformal field theory \cite{CF}. \ The operator $D$
encodes the metric, differential calculus, integration and dynamics. For
simple noncommutative spaces auch as the noncommutative space defined by the
standard model all information about the bosonic and fermionic action is
encoded in the spectrun of the Dirac operator. This is known as the spectral
action principle \cite{CC}. The difficulty in this approach is that in order
to make progress one must know the Dirac operator. Enough information must
be available about $D$ to define geometrical quantities. In the problem at
hand it is not easy to guess what $D$ should one start with. The strategy I\
will adopt is to first gather information about noncommutative spaces with
constant background $B$-fields.

Open strings or D-branes in presence of constant background $B$-field can be
realized by deforming the algebra of functions on the classical world
volume. The operator product expansion for vertex operators is identified
with the star (Moyal) product of functions on noncommutative spaces (\cite{H}%
,\cite{FFZ}). In this respect it was shown that noncommutative U(N)
Yang-Mills theory does arise in string theory.

The star product is defined by
\[
f\left( x\right) \ast g\left( x\right) =e^{\frac{i}{2}\theta ^{\mu \nu }%
\frac{\partial }{\partial \zeta ^{\mu }}\frac{\partial }{\partial \eta ^{\nu
}}}f\left( x+\zeta \right) g\left( x+\eta \right) \left| _{\zeta =\eta
=0}\right.
\]
This definition forces the gauge fields to become complex. Indeed the
noncommutative Yang-Mills action is invariant under the gauge
transformations
\[
A_{\mu }^{g}=g\ast A_{\mu }\ast g_{\ast }^{-1}-\partial _{\mu }g\ast g_{\ast
}^{-1}
\]
where $g_{\ast }^{-1}$is the inverse of $g$ with respect to the star
product:
\[
g\ast g_{\ast }^{-1}=g_{\ast }^{-1}\ast g=1
\]
The contributions of the terms $i\theta ^{\mu \nu }$ in the star product
forces the gauge fields to be complex. Only conditions such as $A_{\mu
}^{\dagger }=-A_{\mu }$ could be preserved under gauge transformations
provided that $g$ is unitary: $g^{\dagger }\ast g=g\ast g^{\dagger }=1.$ It
is not possible to restrict $A_{\mu }$ to be real or imaginary to get the
orthogonal or symplectic gauge groups as these properties are not preserved
by the star product (\cite{SW},\cite{MW}). For open strings in constant
background $B$-field the effective metric is (\cite{CLNY},\cite{SW})
\begin{eqnarray*}
g^{\mu \nu } &=&\left( G_{\mu \nu }+2\pi \alpha ^{\prime }B_{\mu \nu
}\right) _{S}^{-1} \\
\theta ^{\mu \nu } &=&\left( G_{\mu \nu }+2\pi \alpha ^{\prime }B_{\mu \nu
}\right) _{A}^{-1} \\
g_{\mu \nu } &=&G_{\mu \nu }-\left( 2\pi \alpha ^{\prime }\right) ^{2}\left(
BG^{-1}B\right) _{\mu \nu }
\end{eqnarray*}
One can imagine a general setting where the closed string theory metric
arise as an effective metric coming from open strings, or where the D-branes
\ become dynamical. Under such circumstances one can get an effective metric
of the form
\[
g_{\mu \nu }=e_{\mu }^{a}\ast e_{\nu a}
\]
Because of $\theta $ contributions the metric must become complex. This also
seems inevitable as the star product appears in the operator product
expansion of the string vertex operators. We are therefore led to
investigate whether the metric can beome complex.

Assume that we start with the $U(1,D-1)$ gauge fields $\omega _{\mu
\,\,b}^{\,a}$. The $U(1,D-1)$ group of transformations is defined as the set
of matrix transformations leaving the quadratic form
\[
\left( Z^{a}\right) ^{\dagger }\eta _{b}^{a}Z^{b}
\]
invariant, where $Z^{a}$ are $D$ complex fields and
\[
\eta _{b}^{a}=diag\left( -1,1,\cdots ,1\right)
\]
with $D-1$ positive entries. The gauge fields $\omega _{\mu \,\,b}^{\,a}$
must then satisfy the condition
\[
\left( \omega _{\mu \,\,b}^{\,a}\right) ^{\dagger }=-\eta _{c}^{b}\omega
_{\mu \,\,d}^{\,c}\eta _{a}^{d}
\]
The curvature associated with this gauge field is
\[
R_{\mu \nu \,\,b}^{\quad a}=\partial _{\mu }\omega _{\nu
\,\,b}^{\,a}-\partial _{\nu }\omega _{\mu \,\,b}^{\,a}+\omega _{\mu
\,\,c}^{\,a}\omega _{\nu \,\,b}^{\,c}-\omega _{\nu \,\,c}^{\,a}\omega _{\mu
\,\,b}^{\,c}
\]
Under gauge transformations we have
\[
\widetilde{\omega }_{\mu \,\,b}^{\,a}=M_{c}^{a}\omega _{\mu
\,\,d}^{\,c}M_{b}^{-1d}-M_{c}^{a}\partial _{\mu }M_{b}^{-1c}
\]
where the matrices $M$ are subject to the condition:
\[
\left( M_{c}^{a}\right) ^{\dagger }\eta _{b}^{a}M_{d}^{b}=\eta _{d}^{c}
\]
The curvature then transforms as
\[
\widetilde{R}_{\mu \nu \,\,b}^{\quad a}=M_{c}^{a}R_{\mu \nu \,\,d}^{\quad
c}M_{b}^{-1d}
\]
Next we introduce the complex vielbein $e_{\mu }^{a}$ and its inverse $%
e_{a}^{\mu }$ defined by
\[
e_{a}^{\nu }e_{\mu }^{a}=\delta _{\mu }^{\nu },\quad e_{\nu }^{a}e_{b}^{\nu
}=\delta _{b}^{a}
\]
which transform as
\[
\widetilde{e}_{\mu }^{a}=M_{b}^{a}e_{\mu }^{b},\quad \widetilde{e}_{a}^{\mu
}=\widetilde{e}_{b}^{\mu }M_{a}^{-1b}
\]
It is also useful to define the complex conjugates
\[
e_{\mu a}\equiv \left( e_{\mu }^{a}\right) ^{\dagger },\quad e^{\mu a}\equiv
\left( e_{a}^{\mu }\right) ^{\dagger }
\]
With this, it is not difficult to see that
\[
e_{a}^{\mu }R_{\mu \nu \,\,b}^{\quad a}\eta _{c}^{b}e^{\nu c}
\]
is hermitian and $U(1,D-1)$ invariant. The metric is defined by
\[
g_{\mu \nu }=\left( e_{\mu }^{a}\right) ^{\dagger }\eta _{b}^{a}e_{\nu }^{b}
\]
satisfy the property $g_{\mu \nu }^{\dagger }=g_{\nu \mu }.$ When the metric
is decomposed into its real and imaginary parts:
\[
g_{\mu \nu }=G_{\mu \nu }+iB_{\mu \nu }
\]
the hermiticity property then implies the symmetries
\[
G_{\mu \nu }=G_{\nu \mu },\quad B_{\mu \nu }=-B_{\nu \mu }
\]
The gauge invariant Hermitian action is given by
\[
I=\int d^{D}x\sqrt{e}^{\dagger }e_{a}^{\mu }R_{\mu \nu \,\,b}^{\quad a}\eta
_{c}^{b}e^{\nu c}\sqrt{e}
\]
where $e=\det \left( e_{\mu }^{a}\right) .$ One goes to the second order
formalism by integrating out the spin connection and substituting for it its
value in terms of the vielbein. The resulting action depends only on the
fields $g_{\mu \nu }.$ It is worthwhile to stress that the above action,
unlike others proposed to describe nonsymmetric gravity \cite{M} is unique,
except for the measure, and unambiguous. Similar ideas have been proposed in
the past based on gauging the groups $O(D,D)$ \cite{MS} and $GL(D)$ \cite
{Seigel}, in relation to string duality, but the results obtained there are
different from what is presented here. The ordering of the terms in writing
the action is done in a way that generalizes to the noncommutative case. The
idea of a hermitian metric was first forwarded by Einstein and Strauss \cite
{ES}, which resulted in a nonsymmetric action for gravity, with two possible
contractions of the Riemann tensor.

The infinitesimal gauge transformations for $e_{\mu }^{a}$ is $\delta e_{\mu
}^{a}=\Lambda _{b}^{a}e_{\mu }^{b},$ which can be decomposed into real and
imaginary parts by writing $e_{\mu }^{a}=e_{0\mu }^{a}+ie_{1\mu }^{a},$ and $%
\Lambda _{b}^{a}=\Lambda _{0b}^{a}+i\Lambda _{1b}^{a}$ .

From the gauge transformations of $e_{0\mu }^{a}$ and $e_{1\mu }^{a}$ one
can easily show that the gauge parameters $\Lambda _{0b}^{a}$ and $\Lambda
_{1b}^{a}$ can be chosen to make $e_{0\mu a}$ symmetric in $\mu $ and $a$
and $e_{1\mu a}$ antisymmetric in $\mu $ and $a$. This is equivalent to the
statement that the Lagrangian should be completely expressible in terms of $%
G_{\mu \nu }$ and $B_{\mu \nu }$ only, after eliminating $\omega _{\mu
\,\,b}^{\,a}$ through its equations of motion. In reality we have
\begin{eqnarray*}
G_{\mu \nu } &=&e_{0\mu }^{a}e_{0\nu }^{b}\eta _{ab}+e_{1\mu }^{a}e_{1\nu
}^{b}\eta _{ab} \\
B_{\mu \nu } &=&e_{0\mu }^{a}e_{1\nu }^{b}\eta _{ab}-e_{1\mu }^{a}e_{0\nu
}^{b}\eta _{ab}
\end{eqnarray*}
In this special gauge, where we define $g_{0\mu \nu }=e_{0\mu }^{a}e_{0\nu
}^{b}\eta _{ab}$ , $g_{0\mu \nu }g_{0}^{\nu \lambda }=\delta _{\mu
}^{\lambda },$ and use $e_{0\mu }^{a}$ to raise and lower indices we get
\begin{eqnarray*}
B_{\mu \nu } &=&-2e_{1\mu \nu } \\
G_{\mu \nu } &=&g_{0\mu \nu }-\frac{1}{4}B_{\mu \kappa }B_{\lambda \nu
}g_{0}^{\kappa \lambda }
\end{eqnarray*}
The last formula appears in the metric of the effective action in open
string theory \cite{CLNY}.

We can express the Lagrangian in terms of $e_{\mu }^{a}$ only by solving the
$\omega _{\mu \,\,b}^{\,a}$ equations of motion
\begin{eqnarray*}
e_{a}^{\mu }e^{\nu b}\omega _{\nu \,\,b}^{\,c}+e_{b}^{\nu }e^{\mu c}\omega
_{\nu \,\,a}^{\,b}-e^{\mu b}e_{a}^{\nu }\omega _{\nu \,\,b}^{\,c}-e_{b}^{\mu
}e^{\nu c}\omega _{\nu \,\,a}^{\,b} &=& \\
\frac{1}{\sqrt{G}}\partial _{\nu }\left( \sqrt{G}\left( e_{a}^{\nu }e^{\mu
c}-e_{a}^{\mu }e^{\nu c}\right) \right)  &\equiv &X_{\quad a}^{\mu c}
\end{eqnarray*}
where $X_{\quad a}^{\mu c}$ satisfy $\left( X_{\quad a}^{\mu c}\right)
^{\dagger }=-X_{\quad c}^{\mu a}.$ One has to be very careful in working
with a nonsymmetric metric
\[
g_{\mu \nu }=e_{\mu }^{a}e_{\nu a},\quad g^{\mu \nu }=e^{\mu a}e_{\nu
a},\quad g_{\mu \nu }g^{\nu \rho }=\delta _{\mu }^{\rho }
\]
but $g_{\mu \nu }g^{\mu \rho }\neq \delta _{\mu }^{\rho }.$ Care also should
be taken when raising and lowering indices with the metric.

Before solving the $\omega $ equations, we point out that the trace part of $%
\omega _{\mu \,\,b}^{\,a}$ (corresponding to the $U(1)$ part in $U(D)$) must
decouple from the other gauge fields. It is thus undetermined and decouples
from the Lagrangian after substituting its equation of motion. It imposes a
condition on the $e_{\mu }^{a}$%
\[
\frac{1}{\sqrt{G}}\partial _{\nu }\left( \sqrt{G}\left( e_{a}^{\nu }e^{\mu
a}-e_{a}^{\mu }e^{\nu a}\right) \right) \equiv X_{\quad a}^{\mu a}=0
\]
We can therefore assume, without any loss in generality, that $\omega _{\mu
\,\,b}^{\,a}$ is traceless $\left( \omega _{\mu \,\,a}^{\,a}=0\right) .$

The $\omega -$equation gives
\[
\omega _{\kappa \rho }^{\quad \mu }+\omega _{\rho \,\,\kappa }^{\,\,\mu }=%
\frac{1}{8}\delta _{\kappa }^{\mu }\left( 3X_{\,\,\rho \mu }^{\mu
}-X_{\,\,\mu \rho }^{\mu }\right) +\frac{1}{8}\delta _{\rho }^{\mu }\left(
-X_{\,\,\kappa \mu }^{\mu }+3X_{\,\,\mu \kappa }^{\mu }\right) -X_{\,\,\rho
\kappa }^{\mu }\equiv Y_{\,\,\rho \kappa }^{\mu }
\]
We can rewrite this equation after contracting with $e_{\mu c}e_{\sigma }^{c}
$ to get
\[
\omega _{\kappa \rho \sigma }+e_{a}^{\mu }e_{\mu c}e_{\sigma }^{c}\omega
_{\rho \,\,\kappa }^{\,\,\,a}=g_{\sigma \mu }Y_{\,\,\rho \kappa }^{\mu
}\equiv Y_{\sigma \rho \kappa }
\]
By writing $\omega _{\rho \,\,\kappa }^{\,\,\,a}=\omega _{\rho \nu \kappa
}e^{\nu a}$ we get
\[
\left( \delta _{\kappa }^{\alpha }\delta _{\rho }^{\beta }\delta _{\sigma
}^{\gamma }+g^{\beta \mu }g_{\sigma \mu }\delta _{\rho }^{\alpha }\delta
_{\kappa }^{\gamma }\right) \omega _{\alpha \beta \gamma }=Y_{\sigma \rho
\kappa }
\]
To solve this equation we have to invert the tensor
\[
M_{\kappa \rho \sigma }^{\alpha \beta \gamma }=\delta _{\kappa }^{\alpha
}\delta _{\rho }^{\beta }\delta _{\sigma }^{\gamma }+g^{\beta \mu }g_{\sigma
\mu }\delta _{\rho }^{\alpha }\delta _{\kappa }^{\gamma }
\]
In the conventional case when all fields are real, the metric $g_{\mu \nu }$
is symmetric and $g^{\beta \mu }g_{\sigma \mu }=\delta _{\sigma }^{\beta }$
so that the inverse of $M_{\kappa \rho \sigma }^{\alpha \beta \gamma }$ is
simple. In the present case, because of the nonsymmetry of $g_{\mu \nu }$
this is fairly complicated and could only be solved by a perturbative
expansion. Writing $g_{\mu \nu }=G_{\mu \nu }+iB_{\mu \nu }$, and  defining $%
G^{\mu \nu }G_{\nu \rho }=\delta _{\rho }^{\mu }$ implies that
\begin{eqnarray*}
g^{\mu \alpha }g_{\nu \alpha } &\equiv &\delta _{\nu }^{\mu }+L_{\nu }^{\mu }
\\
L_{\nu }^{\mu } &=&iG^{\mu \rho }B_{\rho \nu }-2G^{\mu \rho }B_{\rho \sigma
}G^{\sigma \alpha }B_{\alpha \nu }+O(B^{3})
\end{eqnarray*}
The inverse of $M_{\kappa \rho \sigma }^{\alpha \beta \gamma }$ defined by
\[
N_{\alpha \beta \gamma }^{\sigma \rho \kappa }M_{\sigma \rho \kappa
}^{\alpha ^{\prime }\beta ^{\prime }\gamma ^{\prime }}=\delta _{\alpha
}^{\alpha ^{\prime }}\delta _{\beta }^{\beta ^{\prime }}\delta _{\gamma
}^{\gamma ^{\prime }}
\]
is evaluated to give
\begin{eqnarray*}
N_{\alpha \beta \gamma }^{\sigma \rho \kappa } &=&\frac{1}{2}\left( \delta
_{\gamma }^{\sigma }\delta _{\beta }^{\rho }\delta _{\alpha }^{\kappa
}+\delta _{\beta }^{\sigma }\delta _{\alpha }^{\rho }\delta _{\gamma
}^{\kappa }-\delta _{\alpha }^{\sigma }\delta _{\gamma }^{\rho }\delta
_{\beta }^{\kappa }\right)  \\
&&-\frac{1}{4}\left( \delta _{\beta }^{\kappa }\delta _{\alpha }^{\sigma
}L_{\gamma }^{\rho }+\delta _{\alpha }^{\kappa }\delta _{\gamma }^{\sigma
}L_{\beta }^{\rho }-\delta _{\gamma }^{\kappa }\delta _{\beta }^{\sigma
}L_{\alpha }^{\rho }\right)  \\
&&+\frac{1}{4}\left( L_{\gamma }^{\kappa }\delta _{\beta }^{\sigma }\delta
_{\alpha }^{\rho }+L_{\beta }^{\kappa }\delta _{\alpha }^{\sigma }\delta
_{\gamma }^{\rho }-L_{\alpha }^{\kappa }\delta _{\gamma }^{\sigma }\delta
_{\beta }^{\rho }\right)  \\
&&-\frac{1}{4}\left( \delta _{\alpha }^{\kappa }L_{\gamma }^{\sigma }\delta
_{\beta }^{\rho }+\delta _{\gamma }^{\kappa }L_{\beta }^{\sigma }\delta
_{\alpha }^{\rho }-\delta _{\beta }^{\kappa }L_{\alpha }^{\sigma }\delta
_{\gamma }^{\rho }\right) +O(L^{2})
\end{eqnarray*}
This enables us to write
\[
\omega _{\alpha \beta \gamma }=N_{\alpha \beta \gamma }^{\sigma \rho \kappa
}Y_{\rho \sigma \kappa }
\]
It is clear that the leading term reproduces the Einstein-Hilbert action
plus contributions proportional to $B_{\mu \nu }$ and higher order terms.
We can check that in the flat approximation for gravity with $G_{\mu \nu }$
taken to be $\delta _{\mu \nu }$, the $B_{\mu \nu }$ field gets the correct
kinetic terms. First we write
\[
e_{\mu }^{a}=\delta _{\mu }^{a}-\frac{i}{2}B_{\mu a},\quad e_{\mu a}=\delta
_{\mu }^{a}+\frac{i}{2}B_{\mu a}
\]
The $\omega _{\mu \,\,a}^{\,\,a}$equation implies the constraint
\[
X_{\,\quad a}^{\mu a}=\partial _{\nu }\left( e_{a}^{\mu }e^{\nu
a}-e_{a}^{\nu }e^{\mu a}\right) =0
\]
This gives the gauge fixing condition $\partial ^{\nu }B_{\mu \nu }=0.$ We
then evaluate
\[
\omega _{\mu \nu \rho }=-\frac{i}{2}\left( \partial _{\mu }B_{\nu \rho
}+\partial _{\nu }B_{\mu \rho }\right)
\]
When the $\omega _{\mu \nu \rho }$ is substituted back into the Lagrangian,
and after integration by parts one gets
\[
L=\omega _{\mu \nu \rho }\omega ^{\nu \rho \mu }-\omega _{\mu
\,\,}^{\,\,\,\mu \rho }\omega _{\nu \rho }^{\quad \nu }=-\frac{1}{4}B_{\mu
\nu }\partial ^{2}B^{\mu \nu }
\]
This is identical to the usual expression $\frac{1}{12}H_{\mu \nu \rho
}H^{\mu \nu \rho },$ where $H_{\mu \nu \rho }=\partial _{\mu }B_{\nu \rho
}+\partial _{\nu }B_{\rho \mu }+\partial _{\rho }B_{\mu \nu }.$  The later
developments of nonsymmetric gravity showed that the occurence of the trace
part of the spin-connection in a linear form would result in the propagation
of ghosts in the field $B_{\mu \nu }$ \cite{DD}. This can be traced to the
fact that there is no gauge symmetry associated with the field $B_{\mu \nu }.
$ For the theory to become consistent one must show that the action above
has an additional gauge symmetry, which generalizes diffeomorphism
invariance to complex diffeomorphism. This would protect the field $B_{\mu
\nu }$ from having non-physical degrees of freedom. It is therefore
essential to identify whether there are additional symmetries present in the
above proposed action. This is presently under investigation.

Having shown that it is possible to formulate a theory of gravity with
nonsymmetric complex metric, based on the idea of gauge invariance of the
group $U(1,D-1)$ it is not difficult to generalize the steps that led us to
the action for complex gravity to spaces where coordinates do not commute,
or equivalently, where the usual products are replaced with star products.

First the gauge fields are subject to the gauge transformations
\[
\widetilde{\omega }_{\mu \,\,b}^{\,a}=M_{c}^{a}\ast \omega _{\mu
\,\,d}^{\,c}\ast M_{\ast b}^{-1d}-M_{c}^{a}\ast \partial _{\mu }M_{\ast
b}^{-1c}
\]
where $M_{\ast a}^{-1b}$ is the inverse of $M_{b}^{a}$ with respect to the
star product. The curvature is now
\[
R_{\mu \nu \,\,b}^{\quad a}=\partial _{\mu }\omega _{\nu
\,\,b}^{\,a}-\partial _{\nu }\omega _{\mu \,\,b}^{\,a}+\omega _{\mu
\,\,c}^{\,a}\ast \omega _{\nu \,\,b}^{\,c}-\omega _{\nu \,\,c}^{\,a}\ast
\omega _{\mu \,\,b}^{\,c}
\]
which transforms according to
\[
\widetilde{R}_{\mu \nu \,\,b}^{\quad a}=M_{c}^{a}\ast R_{\mu \nu
\,\,d}^{\quad c}\ast M_{\ast b}^{-1d}
\]
Next we introduce the vielbeins $e_{\mu }^{a}$ and their inverse defined by
\[
e_{\ast a}^{\nu }\ast e_{\mu }^{a}=\delta _{\mu }^{\nu },\quad e_{\nu
}^{a}\ast e_{\ast b}^{\nu }=\delta _{b}^{a}
\]
which transform to
\[
\widetilde{e}_{\mu }^{a}=M_{b}^{a}\ast e_{\mu }^{b},\quad \widetilde{e}%
_{\ast a}^{\mu }=\widetilde{e}_{b}^{\mu }\ast M_{\ast a}^{-1b}
\]
The complex conjugates for the vielbeins are defined by
\[
e_{\mu a}\equiv \left( e_{\mu }^{a}\right) ^{\dagger },\quad e_{\ast }^{\mu
a}\equiv \left( e_{\ast a}^{\mu }\right) ^{\dagger }
\]
Finally we define the metric $g_{\mu \nu }=\left( e_{\mu }^{a}\right)
^{\dagger }\eta _{b}^{a}\ast e_{\nu }^{b}.$ The $U(1,D-1)$ gauge invariant
Hermitian action is
\[
I=\int d^{D}x\left( \sqrt{e}^{\dagger }\ast e_{\ast a}^{\mu }\ast R_{\mu \nu
\,\,b}^{\quad a}\eta _{c}^{b}\ast e_{\ast }^{\nu c}\ast \sqrt{e}\right)
\]
where $e=\det \left( e_{\mu }^{a}\right) .$ This action differs from the one
considered in the commutative case by higher derivatives terms proportional
to $\theta ^{\mu \nu }$. It would be very interesting to see whether these
terms could be reabsorbed by redefining the field $B_{\mu \nu }$, or whether
the Lagrangian reduces to a function of $G_{\mu \nu }$ and $B_{\mu \nu }$
and their derivatives only.

The connection of this action to the gravity action derived for
noncommutative spaces based on spectral triples (\cite{CFF},\cite{CGF},\cite
{C}) remains to be made. In order to do this one must understand the
structure of Dirac operators for spaces with deformed star products.

\end{document}